\documentstyle[pre,aps,psfig]{revtex}

\begin{document}
 \title{Dynamics of two-particle granular collisions on a surface}
\author{ Benjamin Painter and R. P. Behringer}
\address{Department of Physics and Center for Nonlinear and Complex
Systems, Duke University, Durham, North Carolina 27708-0305}
\date{\today}
\maketitle

\begin{abstract}
We experimentally examine the dynamics of two-particle collisions
occuring on a surface.  We find that in two-particle collisions a
standard coefficient of restitution model may not capture crucial
dynamics of this system.  Instead, for a typical collision, the
particles involved slide relative to the substrate for a substantial
time following the collision; during this time they experience very
high frictional forces. The frictional forces lead to energy losses
that are larger than the losses due to particle inelasticity.  In
addition, momentum can be transfered to the substrate, so that the
momentum of the two particles is not necessarily conserved.  Finally,
we measure the angular momenta of particles immediately following the
collision, and find that angular momentum can be lost to the substrate
following the collision as well.
\end{abstract}


\section{Introduction}
Dry granular systems have generated much interest recently in the
physics and engineering communities, both for fundamental
understanding and for direct applications
\cite{jaeger_96,campbell_90,goldhirsch_93b}.
These systems are important both in nature (e.g. avalanches) and in
industry (e.g. pharmeceuticals and grain elevators).

Particles in such systems are typically considered to interact only
through interparticle collisions, i.e.~repulsive contact forces.
Experiments that can yield quantitative data for velocities, collision
rates and other useful quantities are often performed in two
dimensions.  In order to allow reasonable motion of particles in such
an experiment, the particles must either be free to roll or they must
be levitated, for instance by air flow.  Here, we consider particles
rolling on a smooth flat surface.  We note that there are then two
types of friction that the particles experience when in motion.  The
first, rolling friction, occurs when the particle is moving without
sliding on the substrate; its effect is relatively weak, with a
coefficient of friction on the order of $10^{-3}$\cite{kudrolli_97}.
Rolling friction affects individual particles independently of
collisions; it tends to damp motion slowly over time.  It also affects
the mobility of particles on the surface.  For example, segregation
occurs when particles of differing surface properties are shaken on a
smooth surface\cite{tennakoon_99}.  The second type of friction
affecting particles is sliding friction.  This occurs when the contact
point of the particle and the surface is not instantaneously at rest.
Sliding friction can occur when particles undergoing collisions
experience frictional frustration, i.e.~when it is impossible to
maintain nonsliding contacts between colliding particles and the
substrate.  During a collision the contact force between the particles
is much greater than the force of gravity, so some sliding on the
substrate will occur.  Sliding friction is much more dissipative than
rolling friction, with a coefficient of friction on the order of
$10^{-1}$\cite{kondic_99}.  In the experiments described here, the
sliding interactions with the substrate are the predominant mechanism
for energy loss.  The sliding of particles following a collision
leads to an energy dissipation rate that is $\sim10^2$ times greater
than dissipation from rolling friction.  The time over which particles
slide is typically relatively long, $\sim0.05-0.1{\rm s}$.  Hence, the
effective time over which a collision influences the dynamics of a
particle is much longer than the actual contact time of
$\sim10^{-5}{\rm s}$\cite{kondic_99}. After a pair of particles has
stopped sliding, the momentum of their center of mass (in the lab
frame) need not be the same as the before-collision value.  These
features have tremendous importance on dynamics of systems rolling on
surfaces, but have been relatively little explored experimentally.
However, recent related theoretical and numerical work has been
conducted by Kondic\cite{kondic_99}.

The purpose of this paper is to examine in detail some important
aspects of the dynamics of two-particle collisions which occur on a
surface.  We begin by briefly describing the measurement apparatus
used to follow the particles' motion, and then discuss the surface
effects.

\section{Apparatus}
\label{sec:apparatus}

The particles used were $2.38 mm$ steel balls, which moved on a flat
aluminum surface.  The aluminum was black anodized to improve visual
contrast between the steel spheres and the background.  The apparatus
was illuminated from nearly directly overhead; with this lighting,
each metal sphere produced a single bright spot near its highest point
due to the reflection of the overhead light. In order to track the
centers of individual spheres over time, we used high speed video at
rates of 250 frames per second.  We then used particle tracking
techniques to follow the particles.  We began by finding the positions
of the centers of all particles within a video frame, identified by
the brightest points in the image (the local maxima in the brightness
field).  Although there were some secondary reflections between
neighboring balls, these reflections were much less bright than the
primary reflections, and they could be eliminated easily.  By
following the positions of individual particles from frame to frame,
we obtained trajectories, velocities, and other time-varying
quantities of interest.

\section{Particle-substrate dynamics}
\label{sec:2_particle}

\subsection{Rolling friction}

The simplest effect of motion on a substrate is rolling friction, and
we consider this effect first.  The frictional force from a single
sphere rolling on a substrate is usually modeled by
\begin{equation} 
F_{fr} = \mu_r F_N, \label{eqn:rollfrict}
\end{equation} 
where $F_N$ is the normal force at the sphere-substrate contact and
$\mu_r$ is the coefficient of rolling friction.

\begin{figure}[t]
\center{\parbox{6in}{
\psfig{file=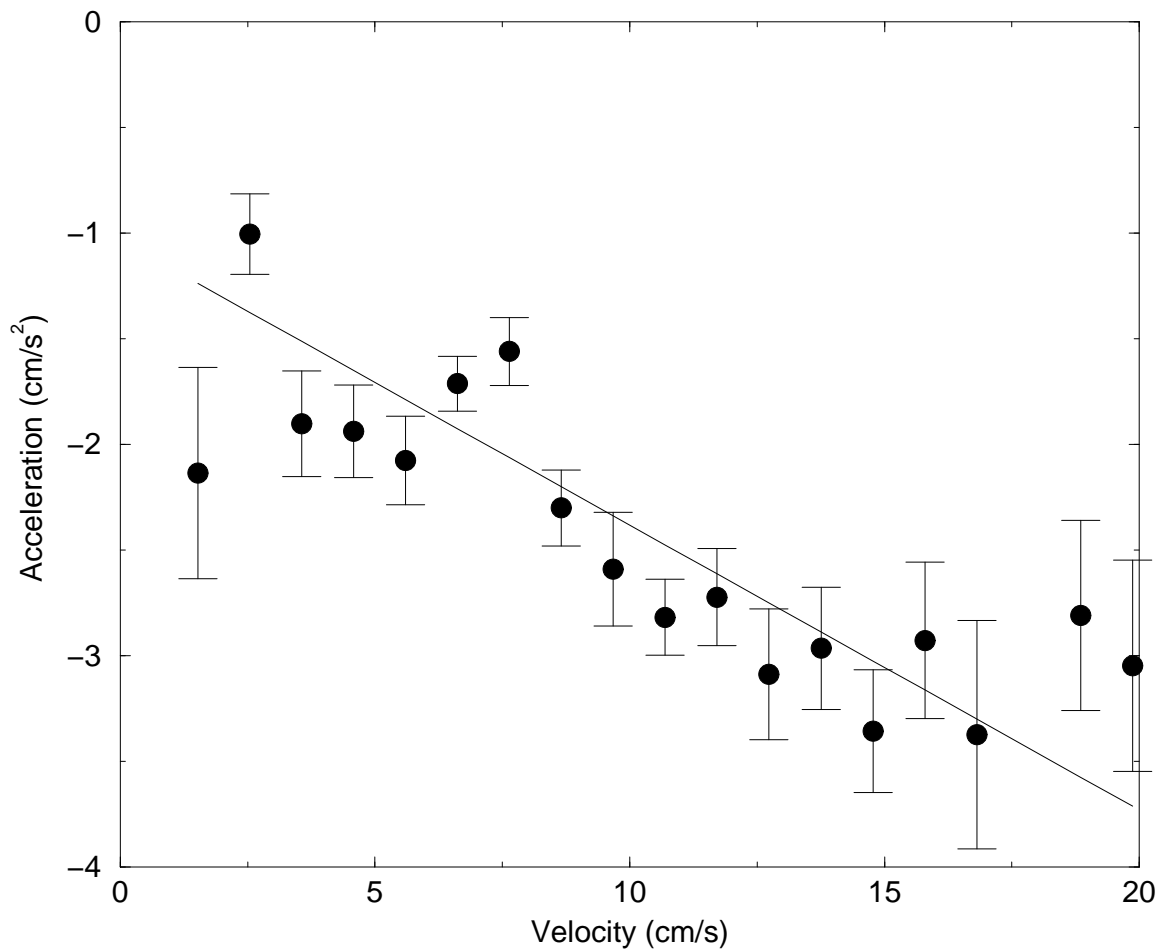,width=6in,angle=270} 
}
\caption{Acceleration due to rolling friction for a single particle
rolling on a horizontal flat surface.  Shown are measured values of
acceleration vs. velocity, and a best-fit line.}
\label{fig:rollfrict}}
\end{figure}

We have carried out measurements of the frictional force on a single
steel sphere rolling on the aluminum substrate described above.  The
sphere was tracked as described in Section \ref{sec:apparatus}.  We
determined its acceleration by dividing the change in velocities
between two frames by the time between the frames; the resulting
acceleration versus velocity is shown in Fig.~\ref{fig:rollfrict}.
The solid line in the figure corresponds to a least-squares linear fit
to the data.  We see that the rolling friction for this system is
velocity dependent, with higher frictional force at higher velocity.
This tends, in principle, to make velocities in rolling granular
systems become more uniform.  To a reasonable approximation, the
acceleration due to rolling friction which a particle experiences is
$a=-C v - D$, where $C=0.135 s^{-1}$ and $D=1.03 cm/s^2$.  Using
$a=\mu_r g$, with a typical acceleration of $a=-2.5 cm/s^2$, we find
that $\mu_r\sim 2.5 \cdot 10^{-3}$, which is comparable to that
reported by Kudrolli $et~al.$\cite{kudrolli_97} for steel balls
rolling on a Delrin surface.

\subsection{Sliding friction}
\label{subsec:slidingfriction}

While this rolling friction has a dissipative effect over long times,
there is another, stronger, mechanism for energy loss to the
substrate: sliding friction.  We find that sliding friction with the
substrate immediately after a collision plays a particularly important
role in the system dynamics.  In order to investigate this effect, we
consider the collision of two particles on a substrate.  We first
review the textbook example of two particles colliding in free space,
which we assume is two-dimensional, and then compare this to
experimental observations when the motion occurs on a substrate.

In the standard case of two inelastic frictionless particles colliding
in free space, i.e.~with no substrate, the collision is described by
conservation of momentum and by an energy loss given through the
coefficient of restitution, $r$.  We introduce the following notation
to describe this process.  The initial momenta of the two particles
are given by $\vec{p}_{1i}$ and $\vec{p}_{2i}$, the final momenta by
$\vec{p}_{1f}$ and $\vec{p}_{2f}$, and $\vec{p}_i =
\vec{p}_{1i}+\vec{p}_{2i}$.  The direction of the vector connecting
the centers of mass of the two particles at the time of the collision
is $\hat{n}$.  The relative velocity of the particles in the $\hat{n}$
direction after colliding is a fraction $r$ of their initial relative
velocity, while the relative velocity tangential to $\hat{n}$ is a
fraction $s$ of its initial value.  Thus,
\begin{eqnarray}
\vec{p}_{1f} + \vec{p}_{2f} & = & \vec{p}_{1i} + \vec{p}_{2i}, \\
p_{1fn} - p_{2fn} & = &-r (p_{1in} - p_{2in}) , \label{eqn:coeffres}
\\ p_{1ft} - p_{2ft} & = &s (p_{1it} - p_{2it}),
\end{eqnarray}
where the subscripts $n$ and $t$ refer to the directions parallel and
perpendicular to $\hat{n}$, respectively.  We take $s=1$, the
simplest case describing an inelastic collision.

This model is usually used in modeling granular
systems\cite{grossman_97,goldhirsch_93a}.  However, it does not
accurately reflect the dynamics of two rolling particles colliding on
a surface.  When two rolling particles collide, there are three
contact points: each particle with the substrate, and the particles
with each other.  In general, these contact points are frictional.
This leads to rotational frustration and, after the collision, to
sliding\cite{kondic_99}.  The following simple argument shows why the
particles are likely to slide on the substrate following a collision.
During a collision the frictional force between the particles competes
with the frictional forces between the particles and the substrate.
If the static friction coefficients at all contacts are comparable,
the frictional force will be greatest where the contact forces are
greatest.  The interparticle contact force is $F_{p-p} \sim \Delta
p/\Delta t$, where $\Delta p = m \Delta v$ is the momentum change of a
particle and $\Delta t_{contact} \approx 10^{-5} {\rm s}$
\cite{kondic_99} is the contact time for a hard-particle collision.
The contact force for a particle with the substrate is $F_{p-s} = m
g$, and the ratio $F_{p-p}/F_{p-s} = \Delta v/(g \Delta t_{contact})$.
Thus $g \Delta t_{contact} \approx 10^{-2} {\rm cm/s}$ for hard metal
spheres defines a crossover velocity, with sliding on the substrate
occuring for $\Delta v > g \Delta t_{contact}$.  If sliding has been
initiated in the collision, a finite time and distance is required
after the spheres separate before dynamic (sliding) friction slows the
spheres' sliding motion.  They will eventually reach a point where
sliding stops and the particles are simply rolling.  During this time,
both the direction and the speed of the particles change
significantly, as detailed below.

\begin{figure}[t]
\center{\parbox{6in}{
\psfig{file=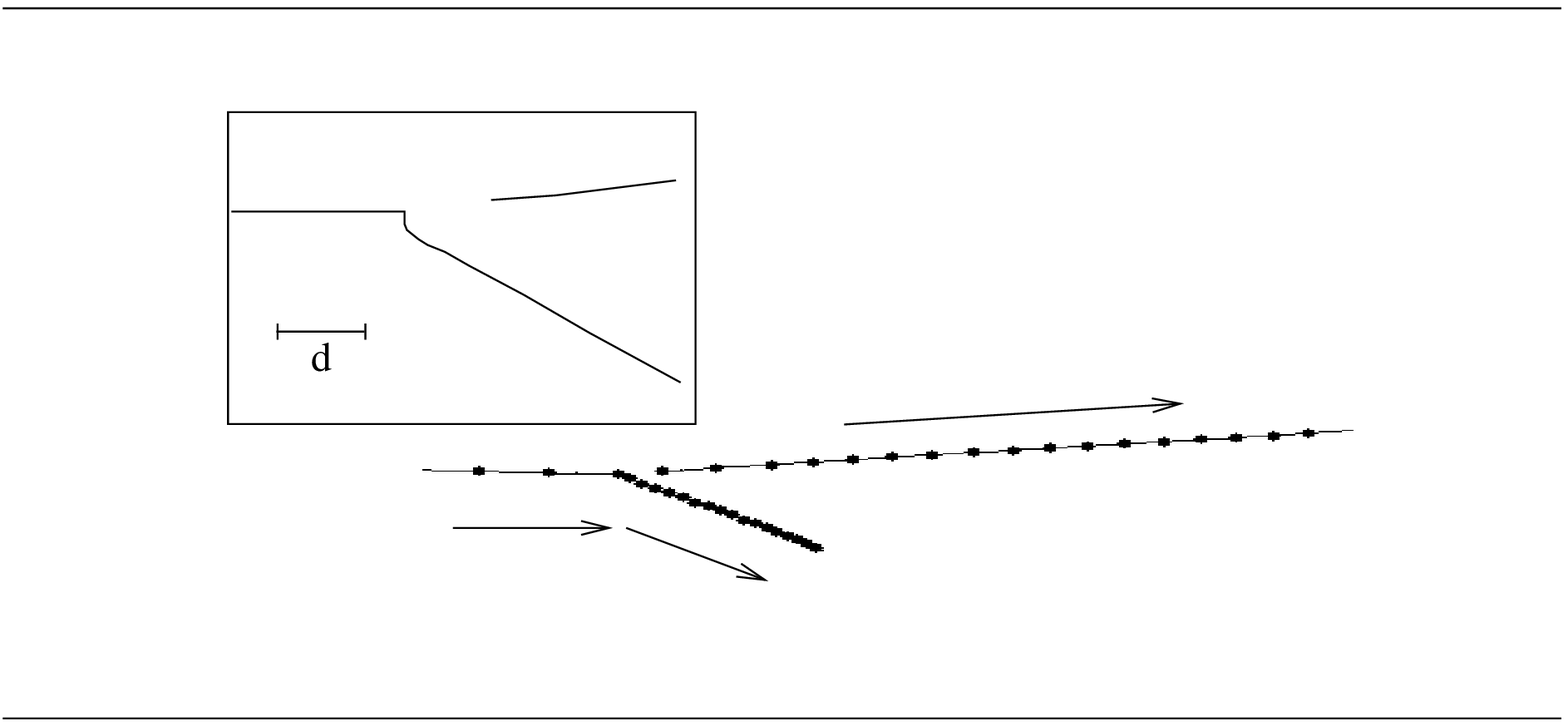,height=3in} 
}
\caption{Tracks of two particles colliding: A moving particle enters
from the left with $v \approx 10 cm/s$ and strikes a stationary particle.
Both exit toward the right.  The circles represent the particles'
positions every $0.02 sec$.  Inset is a detailed view of the particle
tracks near the collision point, with the length of one particle
diameter shown for scale.}
\label{fig:2tracks}}
\end{figure}

We have investigated this effect experimentally by rolling one ball at
an identical stationary ball, and by tracking their motion before and
after the collision.  Fig.~\ref{fig:2tracks} shows a typical set of
trajectories in such a two-particle collision.  The moving ball (in
the lab frame) enters from the left of the image, hitting the
stationary ball.  Since it is difficult to produce a perfectly head-on
collision, the incoming ball strikes the stationary ball slightly
off-center.  Immediately after the collision the two balls behave
almost as though there were no surface interactions (see inset).
Somewhat later, the particles begin to show the influence of the
substrate as they change direction and speed.

\begin{figure}[t]
\center{\parbox{6in}{
\psfig{file=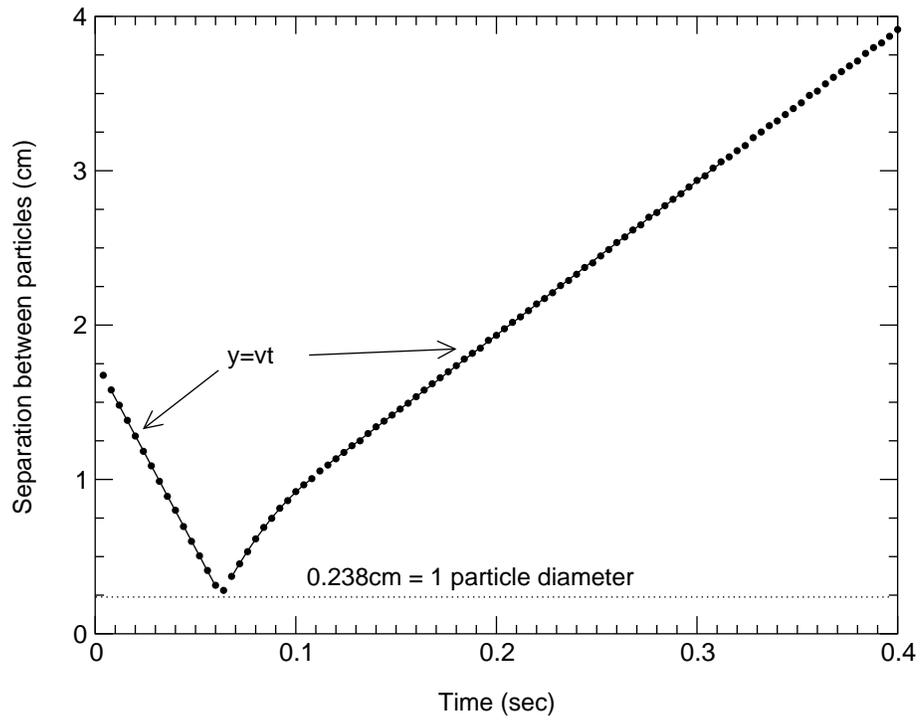,width=6in,angle=270}
}
\caption{Separation $s$ between the centers of the particles shown in 
Fig.~\ref{fig:2tracks} vs. time.}
\label{fig:distapart}}
\end{figure}

During the time between the collision and the time when the balls
begin rolling without sliding, both the direction and the speed of the
balls change due to sliding.  This is shown in
Fig.~\ref{fig:distapart}, which gives the distance between the two
balls shown in Fig.~\ref{fig:2tracks} over time.  The collision occurs
at $t_c \approx 0.06 sec$.  For times $t$ before and well after the
collision, the separation $s$ varies nearly linearly with time,
indicating that the balls roll with nearly constant velocity in these
periods.  By contrast, during the $\sim\,0.06 sec$ immediately
following the collision, the interparticle separation varies
nonlinearly in time.  This indicates a regime in which the two
particles experience dynamic, or sliding, friction with the substrate.
We denote the time following the collision at $t = t_c$ and before the
particles start rolling without sliding at $t = t_r$ as the
``relaxation time,'' $\tau_R = t_r - t_c$.  We define $t_c$ as the
time at which the particles' centers of mass are closest together, and
$t_r$ as the point in time after the collision at which a particle
begins moving with nearly constant velocity.  $\tau_r$ was typically
$0.05-0.1 {\rm s}$ in the systems we studied, which is very large
compared to the time the particles are in contact, roughly
$10^{-5}\,{\rm s}$\cite{kondic_99}.  After a period of time equal to
$\tau_R$ has elapsed, each particle has nearly constant velocity,
affected only by rolling friction.

\begin{figure}[t]
\center{\parbox{6in}{
\psfig{file=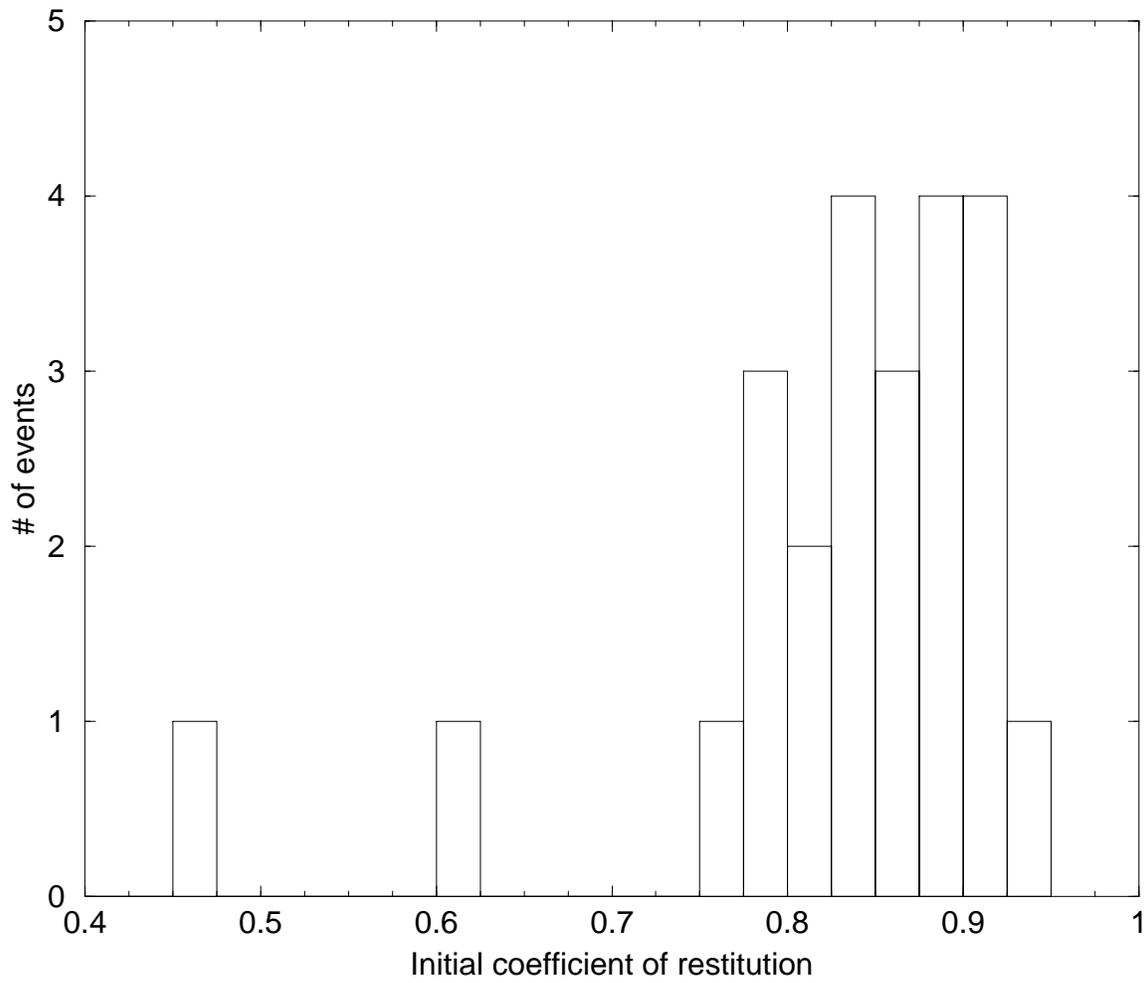,width=6in,angle=270} 
}
\caption{Histogram for the coefficient of restitution, as determined 
immediately following a collision, for 28 samples.}
\label{fig:initcoeff}}
\end{figure}

From collision data we can determine the coefficient of restitution
$r$, as defined in Eq.~\ref{eqn:coeffres} by examining the velocities
immediately preceeding and after the collision, but before sliding
friction has had significant effects.  Fig.~\ref{fig:initcoeff} gives
a histogram of data obtained for a number of measurements of $r$
obtained this way.  To produce these results, we measured velocities
immediately before and within $0.01$ seconds after the collision.  We
find an average value of the coefficient of restitution to be $r =
0.85 \pm 0.11$, with no obvious dependence on the velocity of incoming
particle.  This value is similar to the value of $r$ reported by
others for steel-on-steel collisions: $r = 0.93$ in
ref.~\cite{kudrolli_97} and $r = 0.90$ in ref.~\cite{luding_94}.

\begin{figure}[t]
\center{\parbox{6in}{
\psfig{file=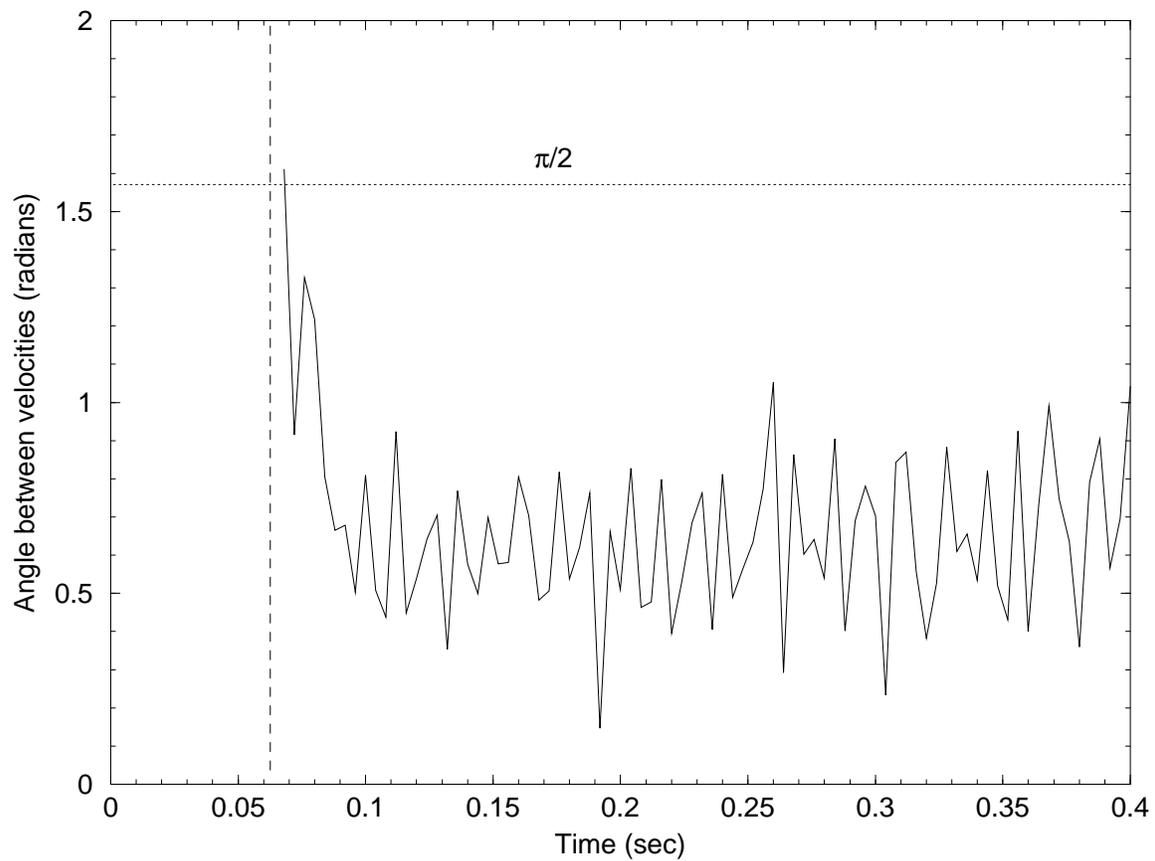,width=6in,angle=270} 
}
\caption{Angle between particle velocities.  The collision occured at
t $\approx 0.06$ sec, as indicated by the vertical dashed line.  At
this point the particles are moving at nearly $90^\circ$ to each
other.  The angle between the incoming particle velocity and $\hat{n}$
is $5.5^\circ$.  The oscillations are due to experimental noise.}
\label{fig:anglebetween}}
\end{figure}

Immediately after the collision, the relative angle of the particles'
new directions is also close to what one would expect for an elastic
collision between two equal-sized spherical particles with one
initially at rest.  Fig.~\ref{fig:anglebetween} shows a typical
example.  In a collision between two identical spheres of radius $R$,
with coefficient of restitution $r$ and impact parameter $b$, the
angle between the directions of the spheres' motion after the
collision is given by
\begin{equation}
cos \theta_f = {{(1-r)(1-(\frac{b}{2R})^2)}\over
{\left\{ \left[ (1-r)^2(1-(\frac{b}{2R})^2)+4(\frac{b}{2R})^2 \right] 
(1-(\frac{b}{2R})^2) \right\}^{1/2}}}.
\end{equation}
Then as $r \rightarrow 1$, a perfectly elastic collision, $cos
\theta_f = 0$ and $\theta_f = \pi/2$, provided that $b/2R \gg 1-r$.  A
typical value of the impact parameter in these experiments is $b/2R
\sim 0.25$.  The collision in Fig.~\ref{fig:anglebetween} occurs at
time $t \approx 0.06 sec$, indicated by the vertical dotted line; at
this time, the relative angle between the particle velcities is near
$\pi/2$.  As sliding friction begins to affect the particles and they
are accelerated or decelerated, the angle between the velocities
decreases.  The direction of the acceleration is discussed in detail
below, in Sec.~\ref{subsec:momentumloss}.

Kondic has investigated a model for two particles colliding on a
surface that includes both the interaction between the particles via a
collision and the interaction of the particles with the substrate
through friction \cite{kondic_99}.  For a system consisting of a
moving particle hitting a stationary particle head-on, he predicts
velocities and relaxation times of each particle after the collision.
If the initial velocity of the moving particle is $v_o$, the final
(i.e.~purely rolling) velocities of the initially stationary and
initially moving particles are $v_{1f}$ and $v_{2f}$ respectively, and
the relaxation times of the initially stationary and initially moving
particles are $\tau_1$ and $\tau_2$ respectively, then:
\begin{eqnarray}
v_{1f} & = & {v_o \over {2(1+{m R^2 \over I})}}((1+r){{m R^2}\over{I}}-2 C) \equiv  F(v_o,r,C) \label{eqn:kondic1}\\ 
v_{2f} & = & {v_o \over {2(1+{m R^2 \over
I})}}(2+(1-r){{m R^2}\over{I}}-2 C)  \equiv  G(v_o,r,C),\\
\tau_1 & = & {{{{1+r}\over{2}}+C} \over {(1+{{m R^2}\over{I}})\mu_k g}}v_o  \equiv  H(v_o,r,C,\mu_k),~{\rm and}\\
\tau_2 & = &{{{{1+r}\over{2}}-C}\over{(1+{{m R^2}\over{I}})\mu_k
g}}v_o  \equiv  I(v_o,r,C,\mu_k) \label{eqn:kondic4},
\end{eqnarray}
where $r$ is the coefficient of restitution of the particles, $\mu_k$
is the coefficient of kinetic friction of the particles with the
substrate, $g$ is the acceleration of gravity, and $I$ is the moment
of inertia of the particles ($I=2 m R^2/5$).  $C$ is a measure of the
transfer of angular momentum between the particles during the
collision, such that immediately after the collision 
\begin{eqnarray}
\omega_1 & = & -C ~\omega_o,~{\rm and} \nonumber \\
\omega_2 & = & (1-C)~ \omega_o, \label{eqn:angulartransfer}
\end{eqnarray}
where $\omega_i$, for $i > 0$, is the angular velocity of each
particle and $\omega_o$ is the angular velocity of the incoming
particle before the collision.

\begin{figure}[t]
\center{\parbox{6in}{
\psfig{file=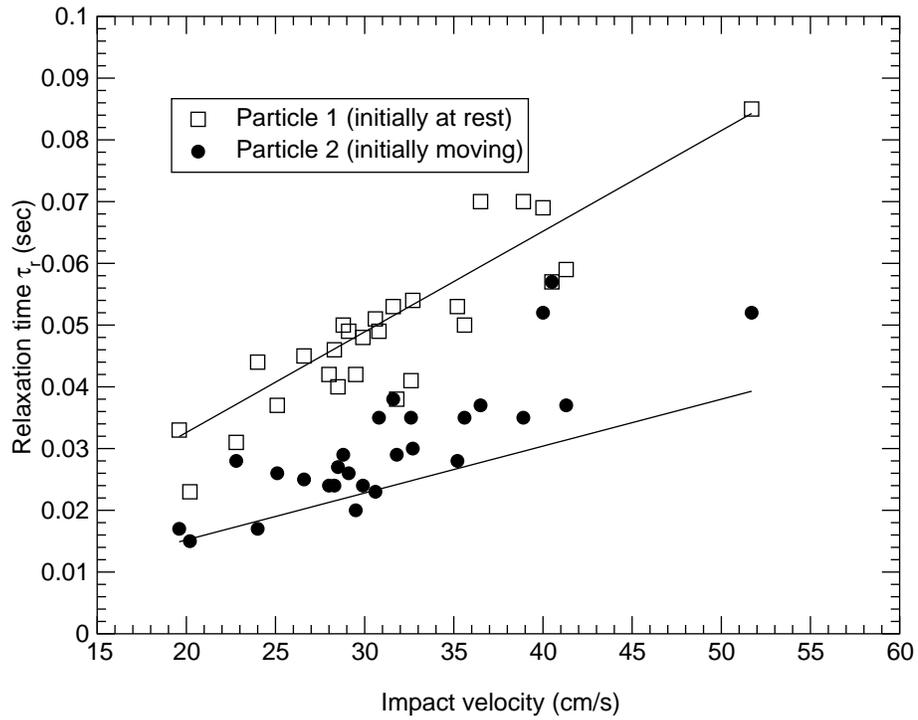,width=6in,angle=270}
}
\caption{Relaxation time vs. impact velocity.  The solid lines are
predictions based on a fit to
Eqs.~\ref{eqn:kondic1}-\ref{eqn:kondic4}.  Note that in a system
without surface interactions, $\tau_r=0$ for all collisions.}
\label{fig:relaxtime}}
\end{figure}

\begin{figure}[t]
\center{\parbox{6in}{
\psfig{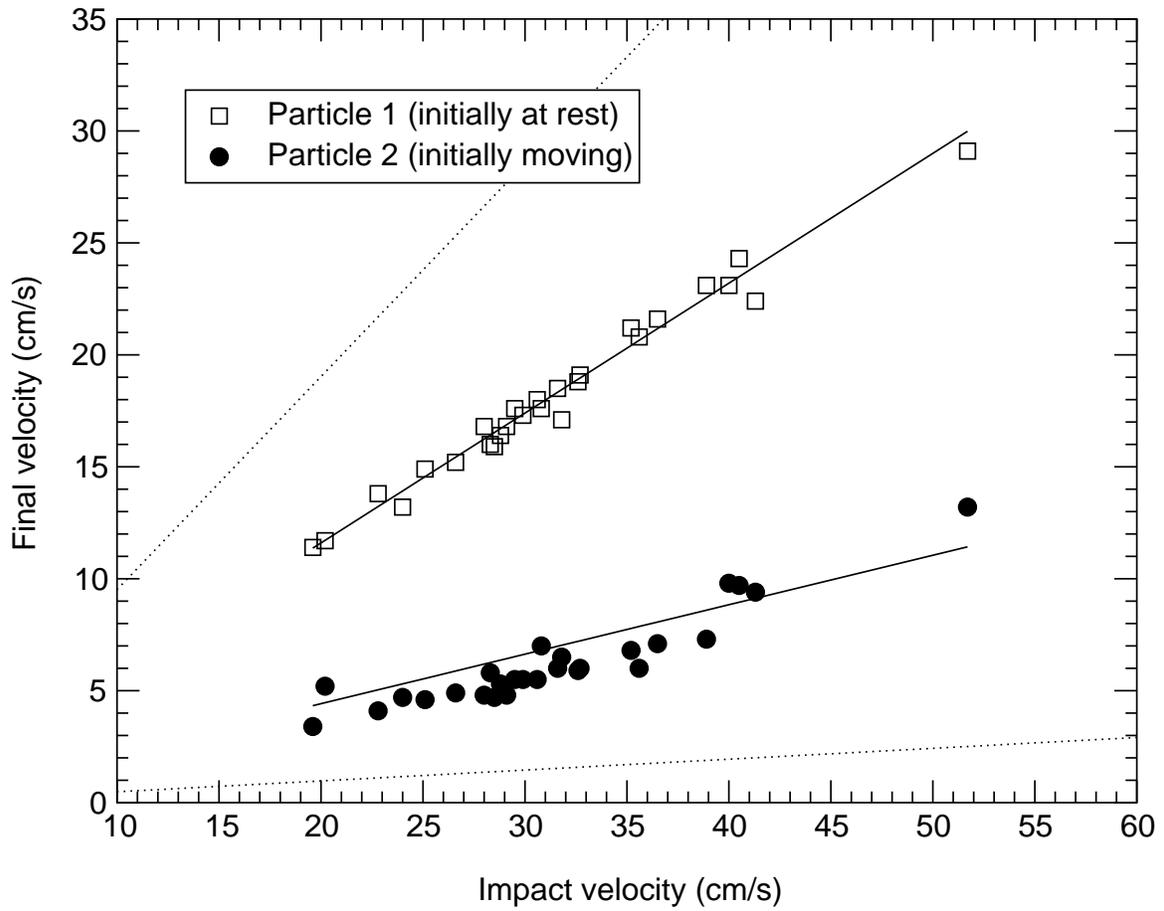}
}
\caption{Final velocity vs. initial velocity.  The solid lines are
predictions based on the fit to
Eqs.~\ref{eqn:kondic1}-\ref{eqn:kondic4}.  Dotted lines represent
theoretical final velocities of particles without surface
interactions, with $r=0.903$.}
\label{fig:relaxvel}}
\end{figure}

We fitted data for experimentally determined final velocities $v_1$
and $v_2$ versus $v_o$ and relaxation times $\tau_1$ and $\tau_2$
versus $v_o$ to these predictions by minimizing the squared deviation
of the model from observed data.  Specifically, we minimized
\begin{eqnarray}
\chi^2 & = {\sum_i} & \left [ {(v_{1fi}-F(v_{oi},r,C))^2
\over{\sigma_{v_{1fi}}^2}} +
{(v_{2fi}-G(v_{oi},r,C))^2 \over{\sigma_{v_{2fi}}^2}} + 
{(\tau_{1i}-H(v_{oi},r,C,\mu_k))^2 
\over{\sigma_{\tau_{1i}}^2}} + \nonumber \right. \\
 & & \left. {(\tau_{2i}-I(v_{oi},r,C,\mu_k))^2 
\over{\sigma_{\tau_{2i}}^2}} \right ],
\end{eqnarray}
with fitting parameters $r$, $\mu_k$, and $C$.  Here $\sigma_n$
represents the experimental uncertainty in the variable $n$.  We found
that $r = 0.903 \pm 0.008$, $\mu_k = 0.232 \pm 0.023$, and $C = 0.347
\pm 0.008$ in our experiments.  Note that this result for the
coefficient of restitution $r$ is consistent with, but much more
precise than, the value of $r=0.85 \pm 0.11$ determined from
Fig.~\ref{fig:initcoeff}.  Results of these fits can be seen in
Fig.~\ref{fig:relaxtime} and Fig.~\ref{fig:relaxvel}.

We conclude that for a two-particle collision on a substrate, the picture of an
instantaneous normal coefficient of friction is inaccurate, and may
not be particularly useful.  Without surface interactions, the
relaxation times are $\tau = 0$ and the final velocities are $v_1/v_o
=(1+r)/2$ and $v_2/v_o=(1-r)/2$ (represented by the dotted lines in
Fig.~\ref{fig:relaxvel}).

\subsection{Energy loss}

\begin{figure}[t]
\center{\parbox{5in} {
\psfig{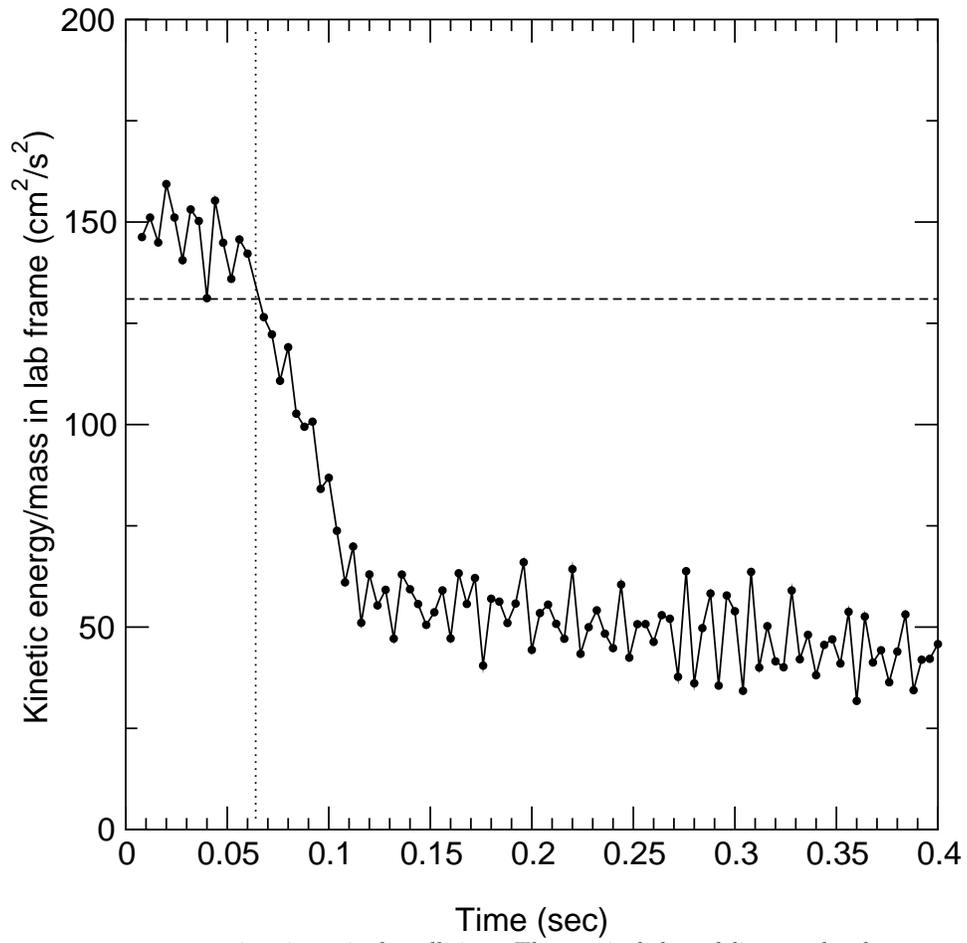} 
}
\caption{Total system energy versus time in a single collision.
The vertical dotted line marks the approximate time of the collision.
The dashed line shows the final energy that would result if the
fractional energy loss was $(1-r^2)/2$.}
\label{fig:energyloss}}
\end{figure}

\begin{figure}[t]
\center{\parbox{6in}{
\psfig{file=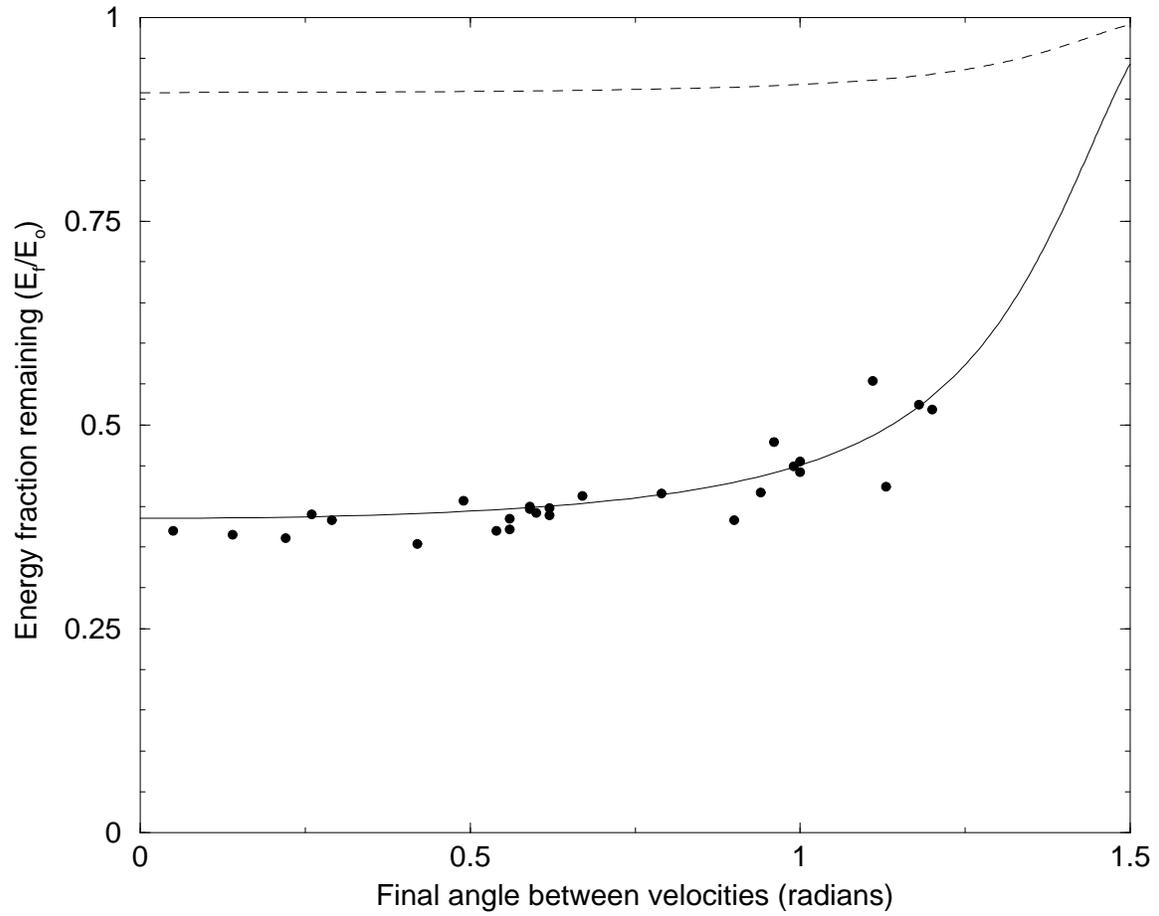,width=6in,angle=270} 
}
\caption{Energy loss vs. angle between final velocities of two particles.
The dashed line represents numerical calculations of two particles
colliding without substrate interactions, with $r=0.903$.  The solid
line represents predictions based on
Eqs.~\ref{eqn:kondic1}-\ref{eqn:kondic4}, with the fitted parameters
$r=0.903$, $\mu_k=0.232$, and $C=0.347$.}
\label{fig:energyvsangle}}
\end{figure}

For many-particle systems, important indicators of the properties of
the collision are the net energy and momentum losses of the system.
Thus we now turn our attention to them.  By the time the particles
have reached the point of rolling without sliding, there has been an
energy loss much greater than that which would occur in a system
described only by a standard coefficient of restitution.  For two
particles undergoing a collision described by a conventional
coefficient of restitution, as in Eq.~\ref{eqn:coeffres}, the
maximum fractional energy loss, which occurs in a head-on collision,
is $(1-r^2)/2$.  For steel balls, with $r \approx 0.903$, $(1- r^2)/2
\approx 0.09$.  For collisions that are not head-on, the energy losses
are smaller, as only the component of velocity normal to the collision
decreases, assuming the tangential coefficient of restitution equals
1.  Figure \ref{fig:energyloss} shows the total system energy versus
time for the two-particle collision described above.  Figure
\ref{fig:energyvsangle} shows the fractional energy remaining in the
system at the end of the relaxation time $\tau_R$ for a series of
collisions as a function of the final angle between the velocities.
For a head-on collision the system's energy after the collision is on
average $\sim 37\%$ of the energy before the collision, representing a
loss of $63\%$ of system energy as the result of a single collision.
Also shown in Fig.~\ref{fig:energyvsangle} is a prediction based on
Eqs.~\ref{eqn:kondic1}-\ref{eqn:kondic4} (solid line) with the
parameters determined in the fit discussed above.  The dashed line
represents energy loss in a system with $r=0.903$ and no surface
interactions. The observed energy loss is only weakly dependent on the
collision angle for nearly head-on collisions.

\subsection{Momentum loss and the direction of sliding frictional forces}
\label{subsec:momentumloss}

\begin{figure}
\center{\parbox{6in}{
\psfig{file=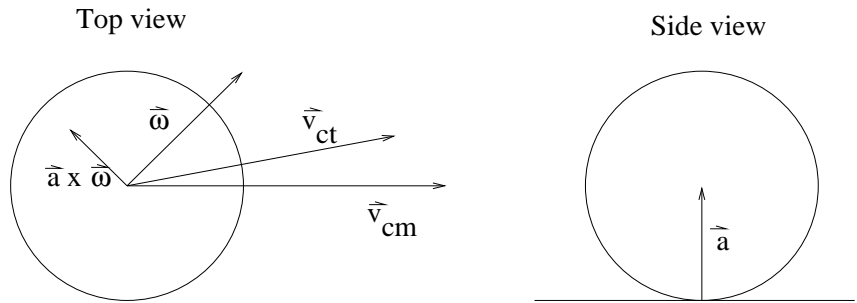,width=6in,angle=270}
}
\caption{Sketch of velocities for a sliding particle.}
\label{fig:veldiagram}}
\end{figure}

\begin{figure}
\center{\parbox{6in}{
\psfig{file=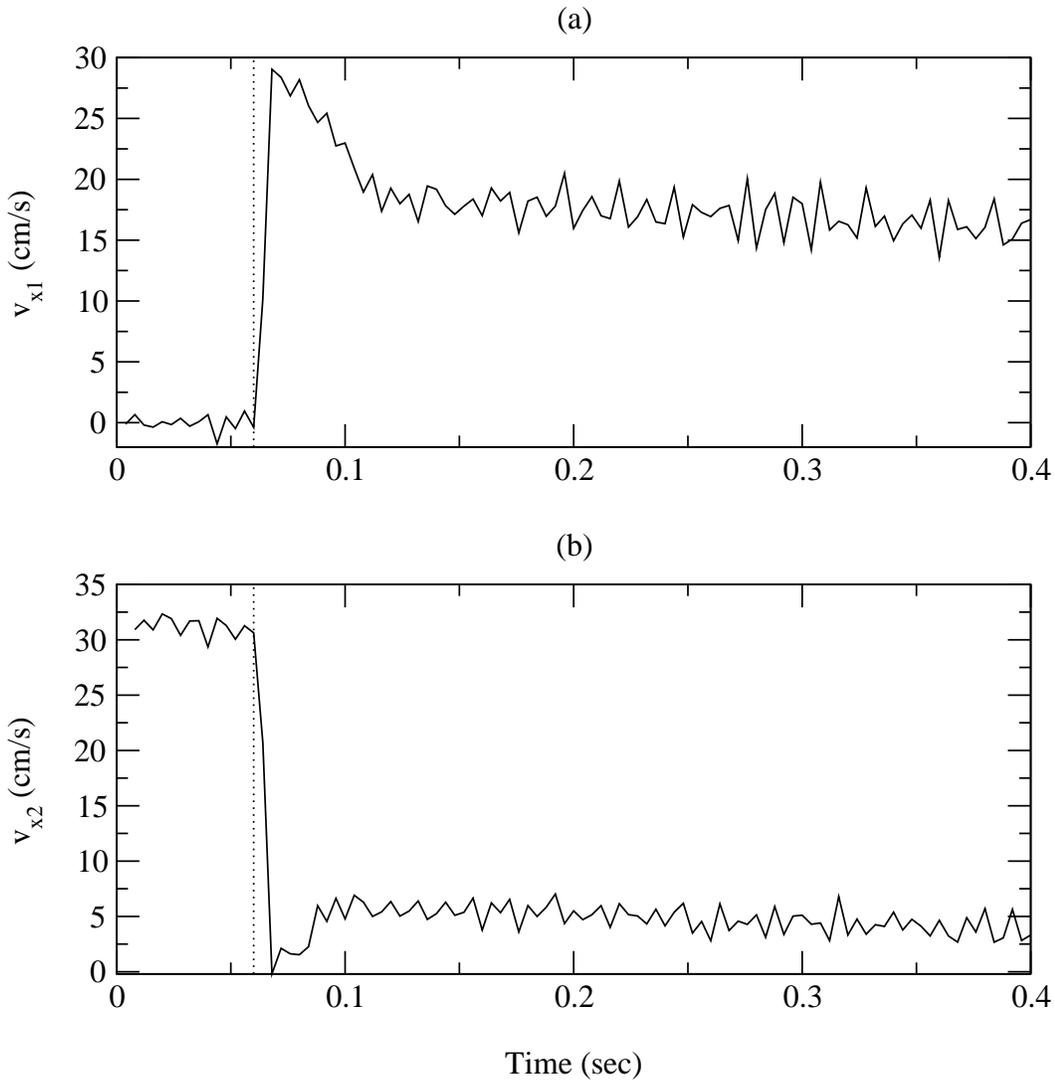,height=6in,angle=270}
}
\caption{Momentum parallel to $\hat{x}$ vs. time, for (a) the
initially stationary particle and (b) the initially moving particle 
($\hat{x}$ is different for each particle).  The collision occurs at $t
\sim 0.06 sec$, as indicated by the vertical dotted lines.  Note that
a large part of the momentum is transfered from one particle to the
other at the time of collision, and that $v_{x2}$ increases after the
collision as a result of its spin.}
\label{fig:pparallel}}
\end{figure}

\begin{figure}
\center{\parbox{6in}{
\psfig{file=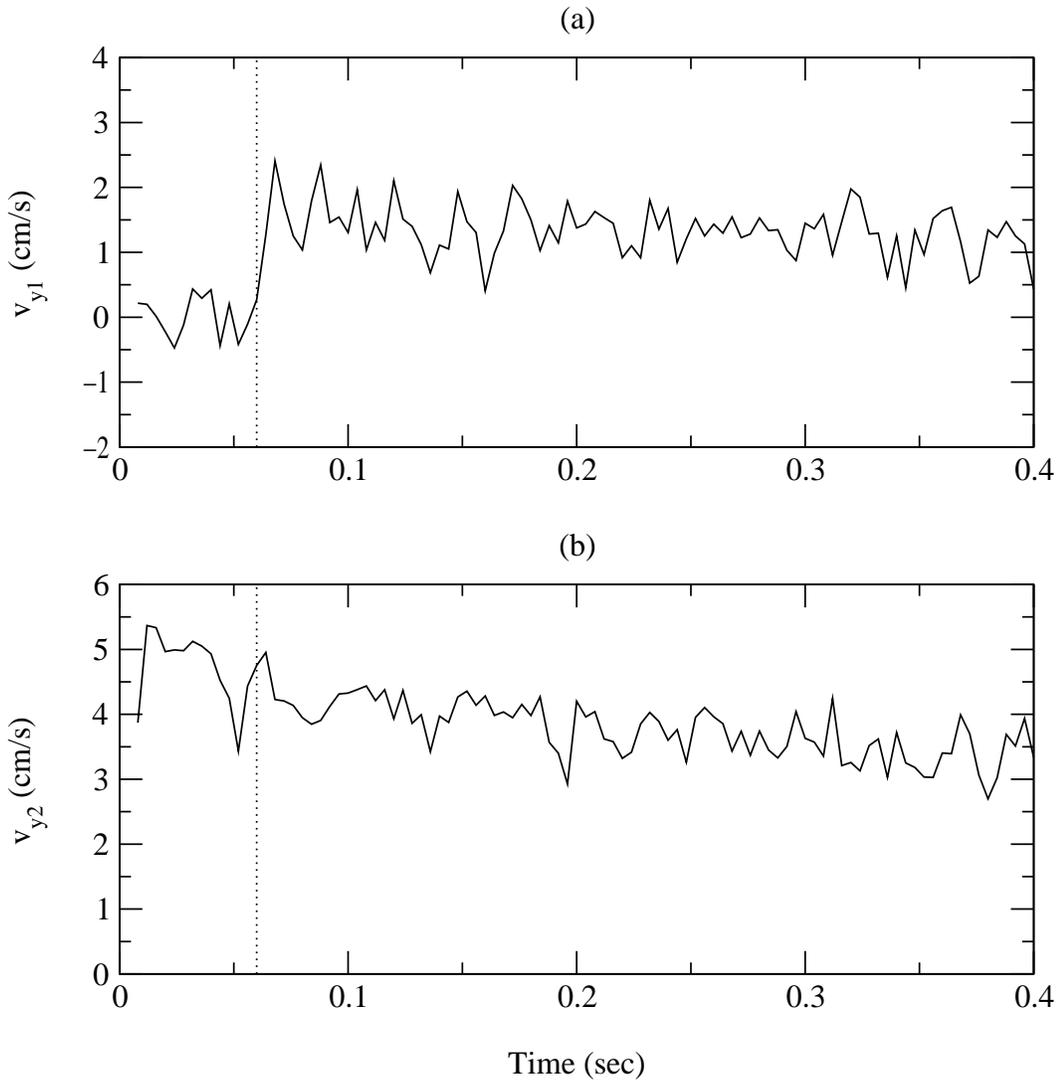,height=6in,angle=270} 
}
\caption{Momentum perpendicular to $\hat{x}$ vs. time, for (a) the
initially stationary particle and (b) the initially moving particle.
The collision occurs at $t \sim 0.06 sec$, represented by the vertical
dotted lines.  Very little momentum transfer takes place in this
direction after the collision.}
\label{fig:pperp}}
\end{figure}

We note that the direction of force due to sliding friction is not
necessarily parallel to the contact normal, $\hat{n}$; instead,
it is in the direction of the relative velocity of the contact between
the substrate and the bottom of the particle, which we call the
contact velocity $v_{ct}$.  Thus,
\begin{eqnarray}
\vec{v_{ct}} = \vec{v_{cm}} + \vec{a} \times \vec{\omega} \label{eqn:vcdef},
\end{eqnarray}
where $\vec{v}_{cm}$ is the velocity of the center of mass of the
particle, $\vec{a}$ is the vector from the contact point to the center
of the particle, and $\vec{\omega}$ is the particle's angular
velocity.  A sketch is provided in Fig.~\ref{fig:veldiagram}.  Here,
we define the $\hat{x}$ direction as $\vec{v_{ct}}/|\vec{v_{ct}}|$,
and the $\hat{y}$ direction as $(\vec{a} \times \hat{x})/|\vec{a}|$.
All momentum loss to the substrate will occur in the $\hat{x}$
direction, as this is the direction of the only force acting on the
particle (neglecting rolling friction, which is small compared to
sliding friction).  We experimentally determine the $\hat{x}$
direction by finding the direction in which a particle's velocity
changes following a collision.  Figures \ref{fig:pparallel} and
\ref{fig:pperp} show the momenta of the individual particles versus
time in the collision described above, in the $\hat{x}$
(Fig.~\ref{fig:pparallel}) and $\hat{y}$ (Fig.~\ref{fig:pperp})
directions.  Note that the $\hat{x}$ and $\hat{y}$ directions are
independently defined for each of the particles, i.e. $\hat{x}$ for
the initially moving particle is different from $\hat{x}$ for the
initially stationary particle.  Figure \ref{fig:pparallel} illustrates
the finite time after the collision (the collision time $t_c$ is
marked by a vertical dotted line) for which momentum is transferred to
the substrate through sliding friction.  After this time, the only
momentum loss is due to rolling friction.  In Fig.~\ref{fig:pperp} we
observe that no momentum is lost in the $\hat{y}$ direction for either
particle after the collision, aside from a slow loss due to rolling
friction.

\begin{figure}
\center{\parbox{6in}{
\psfig{file=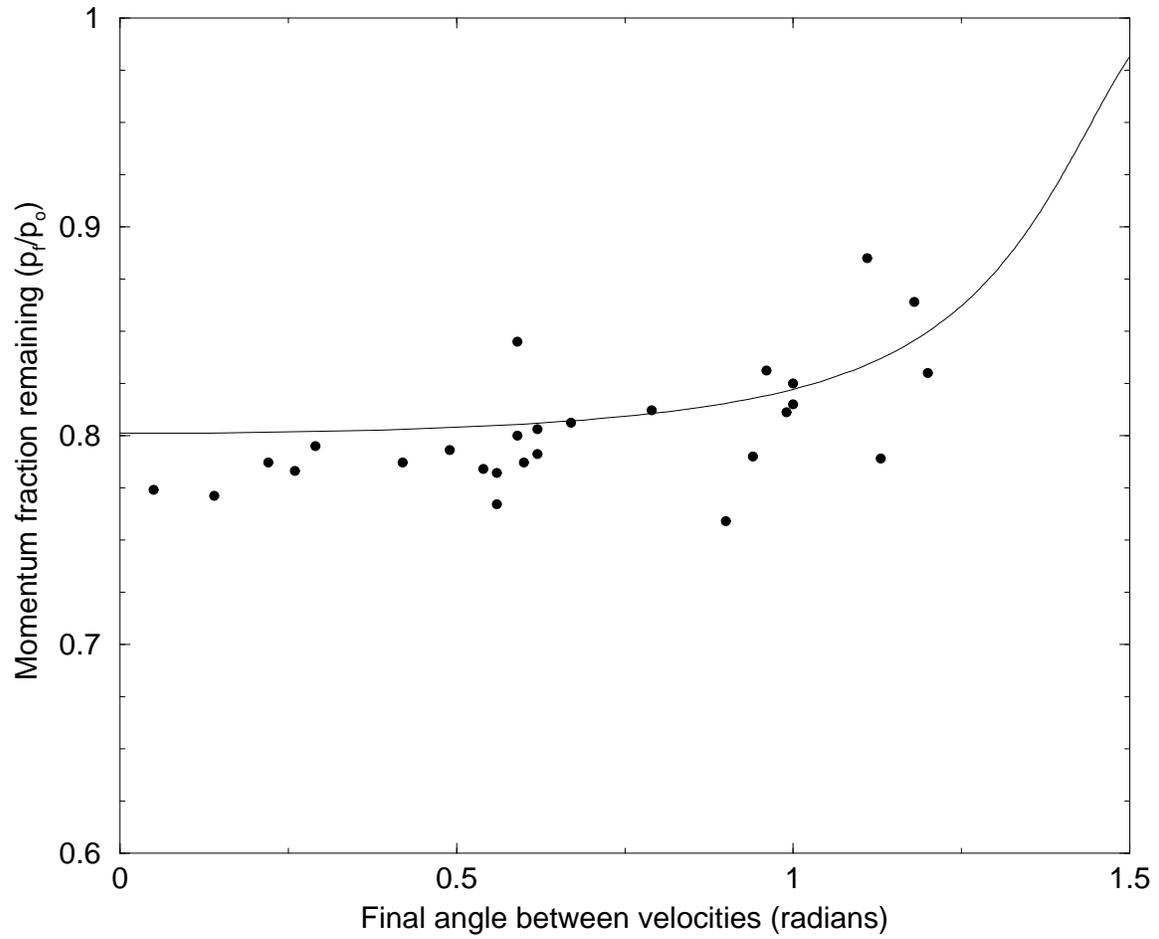,width=6in,angle=270}
}
\caption{Momentum fraction remaining after relaxation (at time 
t=$t_c+\tau_r$) in the direction of the initial momentum (in the lab
frame) versus final angle between velocities.  For a head-on
collision, approximately 20\% of the total system momentum is lost.
The solid line gives the prediction based on
Eqs.~\ref{eqn:kondic1}-\ref{eqn:kondic4}, with the fitted parameters
$r=0.903$, $\mu_k=0.232$, and $C=0.347$.}
\label{fig:pvsangle}}
\end{figure}

We also examine the net momentum loss in the direction of the initial
momentum versus the final angle between the particle velocities
(Fig.~\ref{fig:pvsangle}).  Note that in the usual case, with no
surface interactions, no momentum is lost, so $p/p_0 = 1$ for all
angles.  In contrast, for a head-on collision with surface
interactions we see that $\sim 20\%$ of $\vec{p}_i$ is lost.  Further,
we note that this quantity is weakly dependent on the final angle
between the velocities after the collision for small angles.  The
solid line in this figure shows the prediction based on
Eqs.~\ref{eqn:kondic1}-\ref{eqn:kondic4}, with the parameters
determined above.

This momentum loss may be important in many-particle systems.  For
example, inelastic collapse, a condition in which there are an
infinite number of collisions in a finite time, occurs in one- and
two-dimensional idealized systems \cite{mcnamara_92,mcnamara_94}.
One-dimensional numerical simulations by Dutt $et~al.$ \cite{dutt_99}
show that if even a very small momentum loss per collision is
introduced, inelastic collapse does not occur.  This suggests that
inelastic collapse cannot be observed in experimental granular systems
which interact with a surface.

\section{Angular velocity}

We would also like to determine the angular velocities of the
particles immediately after the collision.  These are difficult to
measure directly, but we can derive expressions for them from
Eq.~\ref{eqn:vcdef}, given the assumption that after sliding stops
each particle will be rolling without sliding.  Then the angular
velocity of a particle immediately after the collision, $\omega_0$, is
\begin{equation}
\omega_{y0} = {1\over{a}}(\nu \Delta v_{cm} + v_{cmx0}),
\end{equation}
and
\begin{equation}
\omega_{x0} = - {1\over{a}} v_{cmy0},
\end{equation}
where $\Delta v_{cm}$ is the change of the center of mass velocity in
the $\hat{x}$ direction due to sliding forces, $v_{cm0}$ is the center
of mass velocity immediately following the collision, and $\nu
\equiv (1+m a^2/I) = 7/2$.  Since we can directly measure the center
of mass velocity at all times, we can deduce the values of
$\omega_{x0}$ and $\omega_{y0}$.

These calculations determine the components of $\vec{\omega}$ in the
$\hat{x}$ and $\hat{y}$ directions, as defined in
Sec.~\ref{subsec:momentumloss}.  These directions are defined by the
sliding frictional forces acting on the particles; a more natural
coordinate system when examining the effect of the collision itself on
angular velocities is defined by the $\hat{n}$ and $\hat{t}$
directions, that is, parallel ($\hat{n}$) and perpendicular
($\hat{t}$) to the vector connecting the centers of mass of the
particles at the time of the collision.  If surface effects are
negligible during the collision, there is no torque in the $\hat{n}$
direction, so we expect that $\omega_{n}$ for each particle will not
be changed by the collision.  Indeed, we find that during the
collision the mean change of the angular velocity in the $\hat{n}$
direction, averaged over 50 particles and normalized by the angular
velocity of the incoming particle in each case, is $d\omega_n /
(v_o/a) = 0.01 \pm 0.02$.  In contrast, we expect that during the
collision, some angular momentum will be transferred from the moving
particle to the stationary particle in the $\hat{t}$ direction.  The
amount of angular velocity transferred can be quantified by
Eq.~\ref{eqn:angulartransfer}; we find from these calculations that
$C=0.25 \pm 0.02$.  This is similar to, although slightly smaller
than, the value of $C=0.347 \pm 0.008$ obtained from the fit to
Eqs.~\ref{eqn:kondic1}-\ref{eqn:kondic4} above.
% in Sec.~\ref{subsec:slidingfriction}.

\section{Conclusion}

In two-dimensional granular systems, understanding interactions with
the substrate is crucial to understanding the system dynamics.  As two
particles collide, there is rotational frustration between them and
the substrate, leading to sliding on the surface.  The large contact
force between the particles at the time of the collision is much
greater than gravity, with $(\Delta v/\Delta t_{coll})/{\rm g} \sim
10^3$, guaranteeing that particles will slide on the substrate after
the collision.  The resulting sliding friction leads to high energy
losses and can be modeled simply, as discussed by Kondic
\cite{kondic_99}. In fact, we find that up to $63\%$ of the incoming
energy is lost in a single collision between two particles with
coefficient of restitution of $0.9$, and most of this is due to
sliding friction with the substrate.  The sliding continues for a
time, $\tau_R$, which is long relative to the collision time.
$\tau_R$ is comparable to or longer than the time between collisions
for moderately dense, rapidly cooling systems, which means that
sliding is experimentally important for many-particle systems until
typical velocities reach $v \sim g \Delta t_{coll}$.  Additionally, we
find that both momentum and angular momentum are typically lost to the
substrate following a collision.

\bibliography{../../../bibtex/granular}

\bibliographystyle{unsrt}

\end{document}